\documentstyle[aps,prd,preprint]{revtex}

\tighten
\begin{document}
\draft

\preprint{\vbox{\baselineskip=12pt
\rightline{gr-qc/9412069}
\rightline{Submitted to Physical Review D}}}

\title{Horizon Boundary Condition for Black Hole Spacetimes}
\author{Peter Anninos$^1$, Greg Daues$^2$, Joan Mass\'o$^{1,3}$,
Edward Seidel$^{1,4}$, and Wai-Mo Suen$^{2}$}

\address{${}^1$National Center for Supercomputing Applications \\
605 E. Springfield Ave., Champaign, Illinois 61820}

\address{${}^2$Department of Physics\\
Washington University, St. Louis, Missouri 63130}

\address{${}^3$Departament de F\'{\i}sica, Universitat de les
Illes Balears, \\   E-07071 Palma de Mallorca, Spain}

\address{${}^4$Department of Physics,
University of Illinois, Urbana, Illinois 61801}

\date{\today}

\maketitle
\begin{abstract}
It was recently shown that spacetime singularities in numerical
relativity could be avoided by excising a region inside the apparent
horizon in numerical evolutions.  In this paper we report on the
details of the implementation of this scheme.  The scheme is based on
using (1)~a horizon locking coordinate which locks the coordinate
system to the geometry, and (2)~a finite differencing scheme which
respects the causal structure of the spacetime.  We show that the
horizon locking coordinate can be affected by a number of shift
conditions, such as a ``distance freezing'' shift,
an ``area freezing'' shift, an ``expansion
freezing'' shift, or the minimal distortion shift.  The causal
differencing scheme is illustrated with the evolution of scalar fields,
and its use in evolving the Einstein equations is studied.
We compare the
results of numerical evolutions with and without the use of this horizon
boundary condition scheme for spherical black hole
spacetimes. With the boundary condition a black hole can be evolved
accurately well beyond $t=1000 M$, where $M$ is the black hole mass.

\end{abstract}


\section{Introduction}
\label{sec:Introduction}

The continued progress in the field of numerical relativity has
demonstrated the feasibility of evolving strong curvature fields,
including black holes, on a computer.  Recent calculations of
spacetimes with black holes include simulations of highly distorted
black holes~\cite{Abrahams92a}, colliding black
holes~\cite{Anninos93b}, the formation of black holes from
imploding gravitational waves~\cite{Abrahams92b}, and balls of
collisionless matter~\cite{Shapiro92a}.  Calculations like these are
important stepping stones to full 3D simulations of two coalescing
black holes.  Such simulations will be very important for
understanding gravitational signals that will be detected by new
gravitational wave detectors, such as the LIGO and
VIRGO~\cite{Abramovici92} laser interferometers and advanced bar
detectors~\cite{Johnson93}.

However, as black holes are accompanied by singularities,
their presence in numerical spacetimes leads to
extreme dynamic ranges in length and time,
making it difficult
to maintain accuracy and stability for long periods of
time.  All calculations of black hole spacetimes to date, like
those mentioned above, develop difficulties at late times due to the large
dynamic ranges that must be computed.  These difficulties are even
more severe when black holes are evolved in 3D~\cite{Anninos94c}.

The traditional way to deal with these problems has been to take
advantage of the coordinate degrees of freedom inherent in the
Einstein equations to avoid the extreme curvature regions.  The ``many
fingers of time'' in relativity allows one to evolve particular regions
of space without evolving the regions in which singularities are
present or forming.  These so-called singularity avoiding slicing
conditions wrap up around the singular region
(see \linebreak
Fig.~\ref{fig:wrap})
so that a large fraction of the spacetime
outside the singular region can be evolved.
Several different types of singularity avoiding slicings
have been proposed~\cite{Smarr78b,Eardley79,Bardeen83} and applied
with variable degrees of success to a number of problems.  However,
these conditions by themselves do not completely solve the problem;
they merely serve to delay the
breakdown of the numerical evolution.
In the vicinity of the singularity,
these slicings inevitably contain a region
of abrupt change near the horizon and a region in which the constant
time slices dip back deep into the past in some sense.  This behavior
typically manifests itself in the form of sharply peaked profiles in
the spatial metric functions~\cite{Smarr78b}, ``grid
stretching''~\cite{Shapiro86}, large coordinate
shift~\cite{Bernstein89} on the black hole throat, {\it etc}.
These features
are most pronounced where the time slices are sharply bent towards the
past (as shown in Fig.~\ref{fig:wrap})
for a reason that will be discussed below.
Numerical simulations will
eventually crash due to these pathological properties of the slicing.
As these problems are even more severe in 3D, where much longer
evolutions will be required to study important problems like the
coalescence of two black holes, it is essential to investigate
alternative methods to handle singularities and black holes in
numerical relativity.

Cosmic censorship suggests that in physical situations,
singularities are hidden inside black hole horizons.
Because the region of spacetime inside the horizon
cannot causally affect
the region of interest outside the horizon,
one is tempted to cut away the
interior region containing the singularity and evolve only the
singularity-free region outside.  To an outside observer no
information will be lost since the region cut away is unobservable.
The procedure of cutting away the singular region
will drastically reduce
the dynamic range, making it easier to maintain accuracy and
stability.  With the singularity removed from the numerical
spacetime, there is in principle no physical reason why black hole
codes cannot be made to run indefinitely without crashing.

Although the desirability of a horizon boundary condition
has been raised many times in the
literature~\cite{Bardeen83,Unruh84,York89},
it has proved to be difficult
to implement such a scheme in a dynamical
evolution~\cite{Thornburg93}.  The boundary condition which one needs to
impose on a black hole horizon, which is a one-way
membrane~\cite{Thorne86}, should be some form of out-going (into the
hole) boundary condition. However, except for the case of linear
nondispersive fields propagating in a flat spacetime, we are not aware
of any satisfactory numerical out-going wave boundary
conditions~\cite{Israeli81}.  Waves in relativity can be nonlinear,
dispersive and possess tails and other complications.  Moreover, what
a wave is in the near zone is not even well-defined.  The development
of a general out-going wave boundary condition in numerical relativity
is certainly highly nontrivial.

In a recent paper~\cite{Seidel92a} we demonstrated that a horizon
boundary condition can be realized. Here we present a more detailed
discussion of our methods and various extensions to that earlier work.
There are two basic ideas behind our implementation
of the inner boundary condition:
(1)~We use a ``horizon locking coordinate'' which locks the
spatial coordinates to the spatial geometry and causal structure.
This amounts to using a shift vector that locks the horizon in place
near a particular coordinate location, and also keeps other coordinate
lines from drifting towards the hole.
In~\cite{Seidel92a} we investigated one particular type of
shift condition, namely the ``distance freezing'' shift.
Here we report on various choices of shift
conditions, including the original ``distance freezing'' shift that
freezes the proper distance to the horizon, an ``expansion
freezing'' shift that freezes the
rate of expansion of outgoing null rays, an
``area freezing'' shift that freezes the area of radial shells, and
the minimal distortion shift~\cite{York79}
that minimizes the global distortion in
the 3-metric.  Some of these shifts have the advantage that
they can be generalized more easily to geometries and coordinate
systems other than the spherical one.  The use of these shift
vectors will be discussed in detail in Section~III.
The basic message is that the idea of a horizon locking
coordinate is robust enough for many different implementations,
with some implementations  likely extendible to the general
3D case.
(2)~We use a finite
differencing scheme which respects the causal structure of the
spacetime, which essentially means that spatial derivatives are
computed at the ``center of the causal past'' of the point being
updated.  Such a differencing scheme is not only essential for the
stability of codes using large shift vectors as in those with
``horizon locking coordinates'', but also eliminates the need of
explicitly imposing boundary conditions on the horizon.
As pointed out in~\cite{Seidel92a},
this is, in a sense,
the horizon boundary condition without a boundary condition.
Since the horizon is a one-way membrane,
quantities on the horizon can be affected only by quantities
outside but not inside the horizon.
Hence, in a finite differencing scheme which respects the causal
structure, all quantities on the horizon can be updated solely in
terms of known quantities residing on or outside the horizon, and
there is no need to impose boundary conditions to account for
information not covered by the numerical evolution.  Such an approach
can be applied to all kinds of source terms that one may
want to deal with in numerical relativity.  [Otherwise one would have
to develop an out-going (into the black hole) wave boundary condition
for each physical problem, i.e., one out-going wave condition for
gravitational waves, one for EM fields, one for perfect fluids,
{\it etc}.]

In Ref.~\cite{Seidel92a} we implemented causal differencing
in a first order way, which respected the causal structure but did not
carefully take account of the exact light cone centers.  In this paper
we report on the results obtained in the second generation of our code
which explicitly and accurately takes account of the causal past of a
grid zone.  The basic idea in constructing the causal differencing
scheme is that, for a set of differential equations written in a
spacetime coordinate system with a shift vector, the finite differenced
version of the equations can be obtained by:$\;$(i)~Transforming
the coordinate system to one without a shift, (ii)~choosing a
differencing method for this set of differential equations in the
usual manner, as required by the physics involved, and
(iii)~transforming
the resulting finite differenced equations back to the original
coordinate system.  The equations so obtained can
be very different from those obtained by applying directly the
usual differencing method
to the set of differential equations with a shift.
A direct application can easily lead to a set of unstable finite
differenced equations when the shift is large.  That is, the actions
of coordinate transformation and finite differencing do not commute.
By going through a coordinate system without a shift vector, we
guarantee that the stability of the final differenced equations is
independent of the shift vector.

A general overview of the program
we adopt is the following: Suppose we want
to numerically evolve a collapsing star. The initial data can be set
up and evolved for a while with some suitable
gauge conditions,
while looking out for the generation of an apparent horizon.  When one
is formed and grows to a certain finite size, a shift vector can be
introduced to maintain the apparent horizon at a constant
coordinate position.  This determines the shift vector right at the
apparent horizon.  The shift at other grid points is determined by
criteria which are consistent
with the choice of the shift at the apparent horizon.
In this paper, we study a subset of the above
scenario in which a spherically symmetric
black hole exists in the initial data. Generalizations of our
basic methods to more complicated geometries are discussed
throughout this paper.

\section{basic equations}
\label{sec:basic}

We use the 3$+$1 formalism which views spacetime as a foliation by
spatial 3-surfaces with a metric $\gamma_{ab}$
and extrinsic curvature tensor
$K_{ab}$. (We adopt the usual notation and use Latin letters
to denote 3-dimensional indices.)  In this picture the spacetime
metric can be written as
\begin{equation}
  ds^2 = \left(-\alpha^2+\beta^a\beta_a\right)dt^2
       + 2\beta_a dx^a dt+ \gamma_{ab}dx^a dx^b,
\label{ds2}
\end{equation}
where $\alpha$ is the lapse function that determines the foliation of
the spacetime and $ \beta^a $ is
the shift vector specifying the three-dimensional
coordinate transformations
from time slice to time slice.

In this formalism, the evolution equations become
\begin{eqnarray}
   \partial_t \gamma_{ab}=
&& -2 \alpha K_{ab} + D_a \beta_b + D_b \beta_a~, \label{evolg} \\
   \partial_t K_{ab}=
&& - D_a D_b \alpha + \alpha \left[ R_{ab} + K K_{ab}
   - 2 K_{ac} K^c_b \right] \nonumber \\
&& +\beta^c D_c K_{ab} +K_{ac} D_b \beta^c + K_{cb} D_a \beta^c,
\label{evolk}
\end{eqnarray}
and the Hamiltonian and momentum constraints are respectively
\begin{eqnarray}
   R + K^2 + K_{ab} K^{ab} = && 0~, \\
   D_a K - D_b{K_a}^b = && 0 ~,
\end{eqnarray}
where $ R_{ab} $ is the 3-Ricci tensor, $R$ is its trace,
$ K $ is the trace of $ K_{ab} $,
and $ D_a $ is the covariant derivative
associated with the 3-metric $ \gamma_{ab} $.

We use the framework
of spherically symmetric spacetimes to
illustrate the idea of locking the horizon and the coordinate system.
As shown by Bernstein, Hobill, and Smarr~\cite{Bernstein89} (denoted
henceforth by BHS), the numerical construction of even a Schwarzschild
spacetime is nontrivial with a general choice of lapse and shift.
Both in this paper and in~\cite{Bernstein89}, the lapse can be arbitrarily
specified.  The shift is taken to be always in the radial direction
with only one component $\beta$, consistent with spherical
symmetry.
The spatial line element is taken to be
\begin{eqnarray}
   d \ell^2 =
&& \gamma_{ab} dx^a dx^b \nonumber \\ =
&& \psi^4 \left( A \, d \eta^2 + B \, d \theta^2
   + B \sin^2 \theta d \phi^2 \right) \, .
\label{3metric}
\end{eqnarray}
The coordinates $ ( \theta , \phi ) $
are the standard spherical coordinates
on the $ \eta = $ constant 2-spheres
and $ \eta $ is a radial coordinate related to the
Schwarzschild isotropic coordinate $ r^{\prime} $ by
$ \eta = \ln (2 r^{\prime} /M) $,
where $M$ is a length scale parameter which is
equal to the mass of the black hole. With this coordinate, the throat
is located at $\eta=0$.  The line element
(\ref{3metric}) is easily generalized to
one which is suitable for numerical studies of axisymmetric
spacetimes~\cite{Abrahams92a}, and it includes both the radial
gauge~\cite{Estabrook73} and the quasi-isotropic or isothermal
gauge~\cite{Shapiro86,Evans86}.
The conformal factor $\psi$
is a function that depends only on $\eta$ and is specified on the
initial time slice so that it satisfies the Hamiltonian constraint
with time symmetry and conformal flatness.

The evolution equations
for the 3-metric and extrinsic curvature (here $\dot{} = \partial_t$
and ${}' = \partial_\eta$) are:
\begin{eqnarray}
   \dot{A} \, = \,
&& \frac{4A \beta \psi '}{\psi } + 2A \, \beta '
   + A' \, \beta  -2\alpha H_A \label{evol_A} \label{adot}\\
   \dot{B} \, =  \,
&& \frac{4B\beta \psi '}{\psi }
   + B' \, \beta -2\alpha H_B \label{evol_B} \label{bdot}\\
   \dot{H}_A =
&& \psi^{-4}\left[ \alpha R_{\eta\eta} - \alpha '' + \alpha '
   \left( \frac{A'}{2A} + \frac{2 \psi '}{\psi }\right)
   + \frac{H_A}{A} \left( 2 \beta ' - \frac{2A' \beta}{A}
   - \frac{4\psi ' \beta }{\psi} \right)
   + \beta \frac{H_A'}{A}\right] \nonumber\\
&& + \, \alpha \, H_A \left( \frac{2 H_A}{A}-\frac{H_B}{B}\right)
   \label{hadot}\\
   \dot{H}_B =
&& \psi^{-4}\left[ \alpha R_{\theta\theta} - \alpha '
   \left( \frac{B'}{2A} + \frac{2 \psi 'B}{A\psi }\right)
   + \frac{4H_B\beta \psi '}{A \psi} + \beta \frac{H_B'}{A}\right]
   +\alpha \frac{H_AH_B}{A}~, \label{hbdot}
\end{eqnarray}
and the Hamiltonian and momentum constraints respectively become
\begin{equation}
  \frac{R_{\eta\eta}}{A\psi^4}
+ \frac{2R_{\theta\theta}}{B\psi^4} + \frac{2H_B^2}{B^2}
+ \frac{4H_AH_B}{AB} = 0~,
\label{H}
\end{equation}
\begin{equation}
   \frac{4H_B\psi '}{B\psi} - \frac{4H_A\psi '}{A\psi}
+ \frac{2H'_B}{B} - \frac{B'H_B}{B^2} - \frac{B'H_A}{AB} = 0 ~,
\label{M}
\end{equation}
where $R_{\eta\eta}$ and $R_{\theta\theta}$ are the 3-Ricci
components and the extrinsic curvature is written as
\begin{equation}
       K_{ab} = \psi^4 \,\, \text{diag} \!
\left( H_A, H_B, H_B \sin^2\theta \right)~,
\end{equation}
to help simplify the form of the equations.

We evolve Eqns.~(\ref{adot}) to (\ref{hbdot})
with the time-symmetric Schwarzschild
solution as the initial data:
$ A = B = 1 $, $ \psi = \sqrt{2M} \cosh(\eta/2)$
and $ H_A = H_B = 0 $.
In the numerically evolved spacetime,
the topology of the $ t = $ constant hypersurfaces is given by the
single Einstein-Rosen bridge, although the geometry can be different.
With the radial coordinate behaving as $\eta \sim \ln r$,
the grid can cover a large range of circumferential radius $r$.  At
the outer boundary, it suffices for our present purpose to take the
metric as fixed.
The inner boundary condition is the subject of this paper.

\section{horizon locking coordinates}
\label{sec:horizon}

\subsection{Motivation}
\label{subsec:motivation}

BHS have developed one of the most accurate codes to date to evolve
single black hole spacetimes in one spatial dimension. They carry out
a thorough treatment using maximal slicing $ ( K = 0 ) $,
with zero or a minimal distortion shift
and nine different methods for finite differencing
the evolution equations including MacCormack, Brailovskaya and
leapfrog schemes. Results presented in this paper will be compared to
this standard.

We first demonstrate the difficulty of using singularity avoiding
slicings (e.g., maximal slicing) using the BHS code.
BHS find that the most accurate
evolutions are obtained by using the
MacCormack or Brailovskaya differencing schemes, maximal
slicing and zero shift vector, although the
leapfrog scheme is of comparable accuracy and is preferred in 2D.
Results obtained with the BHS code
are considered to be very accurate, but as in all codes
designed so far to evolve black holes, it develops difficulties at
late times.
In Figs.~\ref{fig:grr} and~\ref{fig:ham}
we show the best results
using the BHS code with 400 zones to cover the domain
from $\eta=0$ to 6.
The solid lines in Fig.~\ref{fig:grr}
are the radial metric component $ A $ shown from $ t = 0 $ to
$ 100 M $ at every
$ t = 10 M $ intervals. The dashed line is
the coordinate position of the apparent
horizon versus time.  We see that the horizon is growing in
radius, due to the infalling of coordinates and a spike
is rapidly developing near the horizon which eventually causes the
code to crash.  The inaccuracy generated by the sharp spike is shown
explicitly in Fig.~\ref{fig:ham}, where the violation of the
Hamiltonian constraint is plotted at various times.
Figs.~\ref{fig:grr} and~\ref{fig:ham} indicate the code has developed
substantial errors by $ t \sim 90 M$.
At this point, the peak value of the Hamiltonian
constraint stops growing, as the spike in
the radial metric component
can no longer be resolved.
Errors in the apparent horizon mass
are approximately 25\%.
At time $ t \sim 100 M $
numerical instabilities begin
to grow, causing the code to crash shortly thereafter.

The development of the spike is the combined effect of the grid points
falling into the black hole and the collapse of the lapse.  The
coordinate points at smaller radii have larger infall speeds causing
the radial metric component $A$ to increase towards
smaller $\eta$.  However, at the same time, there is a competing
effect due to the use of the singularity avoiding time slice.  The
motion of the grid points close to $\eta =0$ is frozen due to the
``collapse of the lapse.''  At small radii well inside the horizon,
the latter effect dominates, and $A$ cannot increase in
time.  This causes $A$ to develop a peak at a place
slightly inside the horizon where the difference in the infalling
speed of the grid points is large, but the lapse has not completely
collapsed.  BHS investigated using the minimal distortion shift
vector as a means of reducing the shear in the metric
components.  However, they found that although the sharp gradients in
the radial metric component are eliminated from the region containing
the event horizon, the shear is transformed to the throat, where
volume elements vanish as the singularity is approached.

A key to stable and accurate evolutions for long times is to utilize
the shift vector to lock the coordinate system with respect to the
geometry of the spacetime so that there is no infalling of grid
points.  An obvious feature in the black hole geometry that can be
used for such a purpose is the apparent horizon.
Here for convenience of discussion we refer to the
apparent horizon as the two-dimensional
spatial surface having a unit outward pointing 3-vector $ s^{a} $
satisfying\cite{York89}
\begin{equation}
\Theta \equiv D_a s^a + K_{ab} s^a s^b - K = 0 \, ,
\label{theta}
\end{equation}
where $\Theta$ is the expansion of the outgoing null rays.
For a rigorous discussion of the apparent
horizon, see e.g.,~\cite{Hawking73a}.  In this present work, we
assume that there is one and only one such surface in the black hole
spacetime.  Although the apparent horizon is not defined locally in
space, it is defined locally in time,
and hence is a convenient object (in
comparison to the event horizon) to work with in numerical evolutions.

\subsection{Locking the Apparent Horizon}
\label{subsec:horizon}

There are two key steps in locking the coordinate system to the black hole
geometry.  We first lock the position of the apparent horizon to a
fixed coordinate location. Then all grid points
in the spacetime are fixed
with respect to the apparent horizon.

In spherical geometry with the 3-metric (\ref{3metric}) , the
expansion of outgoing null rays reduces to
\begin{eqnarray}
\Theta ( \eta ) &=& \frac{1}{4 \pi \psi^4 B} \, \, \vec\ell
                  \left(4 \pi \psi^4 B \right), \\
       &=& \frac{1}{\psi^2\sqrt{A}}
           \left(\frac{4\psi '}{\psi}+\frac{B'}{B}\right)
           - 2 \, \frac{H_B}{B} \, ,
\label{thetasp}
\end{eqnarray}
where the differential operator
\begin{equation}
\vec\ell = \frac{1}{\alpha}\frac{\partial}
{\partial t} + \left(\frac{1}{\psi^2 \sqrt{A}}
 - \frac{\beta}{\alpha}\right)
    \frac{\partial}{\partial \eta}
\end{equation}
is the outgoing null
vector, the action of which on the surface area is zero at the
apparent horizon.  Therefore the coordinate location of the apparent
horizon $\eta_{AH}$ is given by the root of this equation
$\Theta (\eta =\eta_{AH} )=0$.  One might attempt to solve for
the shift in this equation to make $\eta_{AH}$ constant in time.
However, despite the apparent existence of a $\beta$ term
in the operator $\vec\ell$,
the equation defining the horizon is
independent of $\beta$, once the time
derivative of the metric function $B$ in this equation is expressed in
terms of known quantities on the present time slice.
This is to be expected as the location of the
apparent horizon on a particular time slice should be independent of
the value of $\beta $ on that slice.  Instead, as the time rate of
change of $\eta_{AH}$ is a function of the shift, one can determine
the ``horizon locking'' shift by solving for the shift in the equation
\begin{equation}
\label{shiftcond}
\left.\frac{\partial \Theta(\eta )}{\partial t}
\right|_{\eta = \eta_{AH}} = 0~.
\label{hlock}
\end{equation}
That is, we require the zero of the function $\Theta(\eta )$
at $\eta=\eta_{AH}$ to be
time independent.  This gives a condition on $\beta$.

Alternatively, one can also determine the shift by requiring that the
area of the coordinate surface defining the horizon at some time be
held fixed from that point in time forward.  In a nondynamical
spacetime, such as Schwarzschild, this condition will also lock the
coordinate location of the horizon.  For the present spherically
symmetric case, this requirement can be written as
\begin{equation}
\label{shiftcond2}
\left[ \frac{4\psi'}{\psi}\beta
     + \frac{B'}{B}\beta -2\alpha\frac{H_B}{B}
\right]_{\eta =\eta_{AH}}=0 \, .
\label{arealock}
\end{equation}
One can think of other ways to determine the shift $\beta$ at the
apparent horizon for locking it.  The methods given by
(\ref{shiftcond}) or (\ref{shiftcond2})
are chosen for their extensibility into the 3D case.

Although the present discussion has focused on
``locking'' the horizon, in general one would like to
have the ability to fully
control the motion of the horizon, i.e.,
place the horizon at the coordinate
location of our choosing, which need not be one fixed
value for all times.
For example, if matter is falling into the black hole,
it would be natural to have the horizon of the black hole
expand in coordinate location.
A particularly interesting case is to have the
black hole move across the numerical grid
with the coordinate location of the horizon
changing accordingly in time.
Such controlled motion
of the horizon can be achieved by a simple variant of the
method described above. Work in this direction will be
discussed in detail elsewhere.

\subsection{Shift Conditions}
\label{subsec:coordinate}

Preventing the apparent horizon from drifting is only part
of the story, as we must also specify the shift at other
locations to prevent pathological behavior of the coordinate
system throughout the spacetime and
to prevent grid points from ``crashing'' into
the apparent horizon.  We have investigated the following four
implementations of the shift:

{\bf ``Distance Freezing'' Shift}:\quad  With the apparent
horizon fixed at constant $\eta$, one can tie all grid points
to the horizon by requiring the proper distance between grid
points to be constant in time.
In the spherically symmetric case, this determines the shift
through the differential equation (\ref{adot}).
Setting $ {\dot A} = 0 $ gives
\begin{equation}
\label{shifta}
   \beta^\prime
 + \left( {A^\prime \over 2A} + {2\psi^\prime\over \psi}\right)
  \beta = {\alpha\over A} H_A \, .
\label{dfreeshift}
\end{equation}
Equation~(\ref{shifta}) can be solved
for $\beta$ by integrating from the
horizon to the outer boundary for regions outside the black hole, and
from the horizon to the inner boundary of the numerical grid for
regions inside the horizon.  We use a fourth-order Runge-Kutta method
to solve the first order equation on each time slice.
The use of this shift condition
has been briefly discussed in \cite{Seidel92a}.
In Section~V below, we present
results obtained using this shift, with recent improvements in
implementation incorporated.

{\bf ``Area Freezing'' Shift}:\quad Alternatively, one can
choose to freeze in time the area of
the surfaces of constant
radial coordinate so that ${\dot B} = 0$.
This yields the following equation for $\beta$
\begin{equation}
   \beta
= {2\alpha H_B\psi \over \psi B^\prime + 4B\psi^\prime} \, .
\label{areas}
\end{equation}
An advantage of using the ``area freezing'' shift
is that it is tied nicely
to the apparent horizon when the horizon is locked with
Eq.~(\ref{arealock}).   In fact, Eq.~(\ref{areas}) is simply
an application of Eq.~(\ref{arealock}) not just on the AH, but everywhere.
Another advantage of such a choice is that it yields
an algebraic expression for the shift,
hence eliminating the need for a
spatial integration, as in Eq.~(\ref{dfreeshift}).  Furthermore,
this shift condition allows the surface area
(a sensitive function in the evolution of black hole spacetimes)
to be well-defined in
time and not subject to numerical discretization errors.

However, just like the distance freezing shift, this shift condition
is strongly coordinate dependent.  The usefulness of the distance
freezing shift in a full 3D black hole spacetime depends on one's
ability to pick a suitable ``radial direction'' from the hole, e.g.,
the direction of maximum grid stretching.  Similarly, the usefulness
of the area freezing shift in the full 3D case depends on one's
ability to pick suitable closed 2-surfaces for locking.
In the following we turn to two other shift conditions
that are completely geometric in nature, and therefore can be
generalized in a straightforward manner
to other coordinate systems and to 3D treatment.

{\bf ``Expansion Freezing'' Shift}:\quad   A choice of shift
condition closely related to the area freezing shift
is obtained by freezing the expansion (\ref{theta})
of closed surfaces which have spatially uniform expansions, i.e.,
\begin{equation}
\label{tex}
  \frac{ \partial \Theta }{ \partial t } = 0~.
\end{equation}
In spherical symmetry, these surfaces
are simply surfaces of constant radial coordinates.
We have studied the implementation of both condition~(\ref{tex}) and the
similar condition
\begin{equation}
 \frac{\partial }{ \partial t}
     (\Delta \, \text{Area} ) = \frac{\partial }{ \partial t}
   ( 4 \pi \psi^4 B \Theta ) = 0 \, .
\label{tex2}
\end{equation}
The results are similar.

This shift condition, just like the area freezing shift, ties in
naturally with the horizon locking
condition~(\ref{shiftcond}).
It also yields an algebraic expression for $ \beta $, namely,
\begin{eqnarray}
 \beta &=& \frac{1}{A^2 C' \psi^2} \,
    \left[  4AB \alpha \psi\psi'' + 4 AB \alpha \psi'{}^2
      - 2 \left( A' B + 3 A B' \right) \alpha \psi \psi' \right.
	\nonumber     \\
    &\,& \qquad
  \left.  + \, \left( 2 A^2 R_{\theta \theta} - A C H_A + A B''
     - \textstyle\frac12 A' B' \right)
         \alpha \psi^2  - \sqrt{A} ( A C  \alpha' + A C' \alpha )
	 \right]
\label{tex3}
\end{eqnarray}
with $ C = \psi^4 B \Theta $. On the horizon $ C = 0 $,
but this is not so inside nor outside the horizon.
This condition can be generalized to the 3D case in a straightforward
manner as
the constant $ C $ surfaces can be regarded, in some sense,
as concentric surfaces centered at the
hole. At present we are developing a scheme
for determining such closed 2D
surfaces with uniform expansion in 3D space.

{\bf Minimal Distortion Shift:}
A final option that we consider in this paper
is the minimal distortion shift, written in covariant form
as
\begin{equation}
D^b D_b \beta_a +D^b D_a \beta_b -\frac{2}{3} D_a D_b \beta^b
= D^b \left( 2\alpha \left( K_{ab} - \frac{1}{3} \, \gamma_{ab} \, K
\right)\right)~.
\end{equation}
In spherical coordinates, this reduces to the following
second order equation for $\beta$
\begin{eqnarray}
     \beta '' =
&& -\left( \frac{B'}{B} + \frac{A'}{2A}\right) \beta '
   -\left( \frac{3}{4} \frac{A'B'}{AB} - \frac{B ''}{2B}
   - \frac{B'{}^2}{4 B^2} + \frac{A''}{2A}
   - \frac{ A'{}^2 }{ 2 A^2} \right) \beta \nonumber  \\
&& + \frac{H_B}{B} \alpha' - \frac{H_A}{A} \alpha'
   + \frac{H'_B}{B} \alpha
   + \frac{B' H_B \alpha }{2B^2}
   - \frac{H'_A \alpha }{A} + \frac{A'H_A \alpha}{A^2}
   - \frac{3B'H_A \alpha}{2AB} \,\, .
\label{mini}
\end{eqnarray}
One of the most attractive properties of this shift condition
is its geometric nature. It minimizes coordinate shear in
a global sense. Also its formulation is completely independent
of coordinates so that it may be equally applied to a single
black hole in spherical coordinates as to a two black hole
coalescence in 3D Cartesian coordinates.  However, this shift
vector is more difficult to implement numerically, particularly
in three dimensions.
We solve separately for the two regions inside and outside
the apparent horizon. Outside the horizon we treat the equation
for $\beta$ as a two point boundary value problem using
(\ref{hlock}) or (\ref{arealock})
to fix $\beta$ at the horizon and setting $\beta=0$ at the outer edge.
Inside the horizon
we use a second order backward substitution method
specifying $ \beta $ and $ \beta' $ at the horizon.
$ \beta' $ is computed from the outer domain solution, thus
allowing for smooth extensions of the numerical solution
through the horizon.  The
difficulty in extending this to the 3D case is that one has to
solve a set of coupled elliptic partial differential equations with an
irregular inner boundary on each time slice during the evolution.

All the above discussed shift conditions have been found to
successfully lock the coordinate system to the spacetime
geometry in the spherically symmetric case.  The results will be
given in Section~V.
The basic point we want to make here is that the
idea of locking the coordinate system to the geometry in black hole
spacetimes by making use of the apparent horizon is
robust enough that there can be many different ways to implement
it.  This robustness makes it promising for implementation in 3D.

An important point to note is that these shifts can be applied to
either all of the spacetime grid, or just in the vicinity of the
black hole.  The freedom of turning the shift off at a distance
away from the black hole, so that the coordinates are not necessarily
everywhere locked, is important when we go away from spherical
symmetry.  For example, in multiple black hole spacetimes, one
would like to be able to lock the grid in one part of the
spacetime to one hole, while locking other parts to other black holes.
We will demonstrate in Section~V that such partial locking is
possible.

\section{CAUSAL DIFFERENCING}
\label{sec:causal}

One consequence of introducing a nonzero shift vector in horizon
locking coordinates is that inside the horizon the future light cone is
tilted inward towards smaller $\eta$.  If the shift is such that the
horizon stays at constant coordinate value, i.e., the horizon locking
shift of Section~III above, the light cone will be completely tilted to
one side inside the horizon.
This feature of the light cone is
convenient for the implementation of the horizon boundary condition.
It allows us to excise the singular region
inside some fixed grid point.
Since grid points are fixed to the coordinates,
data at a particular grid point depends only on past data
from grid points at equal or larger {\it coordinate}
values inside the horizon.
Note that without such a horizon locking shift vector this will not
necessarily be true.
The remaining task is then to construct a finite
differencing scheme which can maintain
the causal relations between grid points.
Of course, due to the Courant stability requirement,
causal relations between grid points cannot be
maintained exactly. Inevitably information propagates slightly
faster than the speed of light in a finite differencing
equation. However, by keeping buffer zones
inside the horizon in the numerical evolution domain, the
light cones will be tilted to such an extent that even the innermost
grid point can be evolved with information on grid points
in the buffer zones while
having the Courant condition safely satisfied.
With such a scheme there is no need to supply boundary
conditions at the inner edge of the grid, since all points which can be
affected by the inner edge point, even on the finite differencing
level, are off the grid.

We shall see that the
causal differencing scheme is useful not only
for imposing the horizon boundary condition,
but that it is also essential for stability
when evolving the differential equations
with a large shift vector~\cite{Alcubierre94a}.
In the following
subsections, we shall first illustrate this with a simple
scalar field example,
before turning to the general relativistic case.

\subsection{Scalar Field Example}

Consider a simplified $1\!+\!1$ flat space
\begin{equation}
    ds^2 = - \, d {\tilde t}^{\, 2} + d {\tilde x}^2~,
\label{28}
\end{equation}
and a simple scalar field $\phi$ described by
\begin{equation}
\label{scalev}
  \frac{\partial^2 \phi }{ \partial {\tilde t}^2}
 = \frac{\partial^2 \phi }{ \partial {\tilde x}^2} \, .
\label{29}
\end{equation}
Introduce a shift vector by performing a coordinate transformation
\begin{eqnarray}
 \tilde t \,\,  && = t~,  \\
 \tilde x \,\,  && = f(t,x) \, .
\end{eqnarray}
The spacetime becomes
\begin{equation}
   ds^2 = - \, d t^2 + f'{}^2 (dx + \beta dt)^2
\end{equation}
with
\begin{equation}
   \beta = \frac{ {\dot f} }{ f' } \,\, .
\end{equation}
The evolution equation (\ref{scalev}) in the first order form becomes
\begin{equation}
  {\dot \phi} = K
\label{34}
\end{equation}
\begin{equation}
    {\dot K}
  = 2 \beta K' + \beta' \phi' - \beta^2 \phi''
  - \beta \beta' \phi' - \frac{ \phi'' }{ f'{}^2 }
  - \frac{ f'' \phi' }{ f'{}^2  }   \, .
\label{35}
\end{equation}
One might attempt to finite difference (\ref{34}) and
(\ref{35}) in terms of the usual leapfrog scheme:
\begin{equation}
  {\phi_j^{n+1} - \phi_j^{n-1} \over 2\Delta t} = K_j^n
\label{36}
\end{equation}
\begin{equation}
  {K_j^{n+1} - K_j^{n-1} \over 2\Delta t}
= 2 \beta_j^n \, {K_{j+1}^n - K_{j-1}^n \over 2\Delta t} + \cdots
\label{37}
\end{equation}
in obvious notation. However, a straightforward von Neumann
stability analysis shows that
for any given $ \Delta t / \Delta x  $,
the system of finite difference equations (\ref{36}) and (\ref{37})
is {\it unstable} for a large enough $ \beta $.
Fig.~\ref{fig:cenun} demonstrates the
development of $ \phi$ obtained by this scheme.  The initial data
for $\phi$ is a gaussian
represented by the dashed line and
$ \Delta t / \Delta x = 0.7 $.  The shift
$ \beta $ is taken to be the following function of space and time
\begin{equation}
  \beta = \frac{ 12 ( t + 500 ) }{ 5 (x + 4500) } \, ,
  \label{scalsf}
\end{equation}
so that the field $\phi$ is experiencing
a shift increasing in time, but decreasing in $ x $,
analogous to the black hole case.
$ \phi $ is plotted at equal time intervals
up to time $ t = 210 $ as solid lines in the figure.
The initial gaussian splits into two.
Due to the shift
one component has a large
coordinate velocity moving rapidly to the left, while
the other component has a much smaller coordinate velocity.
We see that at the last plotted
time $ t = 210 $, the evolution becomes unstable.

The reason for this instability is easy to understand.  In trying
to update the data at the $ j^{\text{th}} $ point on the
$ n^{\text{th}} $
slice [the point $ (n+1,j) $],
the ``region of finite differencing'' used on
the right-hand side of (\ref{37})
is from $ (n, j-1) $ to $ (n,j+1) $.  However, the
point $ (n+1,j) $ has the edges of its backward light cone on the
$ n^{\text{th}} $ slice, with
the presence of a shift $ \beta $, given by
$ j - {\Delta t\over\Delta x} + {\beta\Delta t\over\Delta x} $
on the left, and
$ j + {\Delta t\over\Delta x} + {\beta\Delta t\over\Delta x} $
on the right.  Hence if $ \beta + 1 > {\Delta x \over \Delta t}$,
or $ | \beta - 1 | > {\Delta x\over \Delta t} $
for a negative $ \beta $,
part of the ``region of causal dependence'' of the point
$ (n+1,j) $ lies outside the
``region of finite differencing'', and
the evolution becomes unstable.

To enforce that the finite differencing scheme follows the causal
structure of the numerical grid, which becomes nontrivial in the
presence of a shift, the idea we propose is the following:
\begin{itemize}
\item[(i)]
In trying to finite difference a set of differential equations
with a shift, like Eqns.~(\ref{34})
and (\ref{35}), we first transform back to
a coordinate without shift.  In the case above, this means going
back to (\ref{28}).
\item[(ii)]
In this new coordinate system without shift, the causal structure
is trivial, and the finite differencing scheme can be picked in
the usual manner according to the structure of the equation.  In
the present case, for example, it can be the leapfrog scheme
\begin{equation}
  {\phi_j^{n+1} - \phi_j^{n-1} \over 2\Delta \tilde t} = \Pi_j^n
\label{39}
\end{equation}
\begin{equation}
 {\Pi_j^{n+1} - \Pi_j^{n-1} \over 2\Delta \tilde t}
= {\phi_{j+1}^n - 2\phi_j^n + \phi_{j-1}^n
  \over (\Delta \tilde x )^2} \, .
\label{40}
\end{equation}
Notice that we have denoted
$ \displaystyle\frac{ \partial \phi }{ \partial \tilde t } $ as $ \Pi $,
instead of $ K $,
since $ \displaystyle\frac{ \partial \phi }{ \partial \tilde t } $
is different from
$ \displaystyle\frac{ \partial \phi }{ \partial t }. $ \linebreak
\item[(iii)]
The finite differenced equations (\ref{39}) and (\ref{40})
are in terms of
$ \tilde t$ and $\tilde x$.
We then transform these
difference equations back
to the coordinate system $ (t , x) $ with a
shift.  This procedure gives
the ``causal leapfrog differencing''
\begin{equation}
  {\phi_j^{n+1} - \phi_j^{n-1} \over 2\Delta t}
= \Pi_{j+\epsilon}^n + \beta_{j+\epsilon}^n \,
  {\phi_{j+2\epsilon}^{n-1} - \phi_j^{n-1}\over 2 \epsilon \, \Delta x}
\label{41}
\end{equation}
\begin{equation}
  {\Pi_j^{n+1} - \Pi_j^{n-1} \over 2\Delta t}
= {\phi_{j+1+\epsilon}^n-2\phi_{j+\epsilon}
   ^n+\phi_{j-1+\epsilon}^n \over 2\Delta x}
+ \beta_{j+\epsilon}^n \, {\Pi_{j+2\epsilon}^{n-1}
          -\Pi_j^{n-1} \over 2 \epsilon \, \Delta x}
\label{42}
\end{equation}
where $\epsilon = \beta{\Delta t\over \Delta x}$.  Notice that
here the dependent variable $\Pi$ is {\it not$\,$}
transformed back to $K$.
In using the variable $\Pi$, the final equations are simpler and do
not involve the derivative of $ \beta $.
The price to pay is that, if
we want to reconstruct $K$ from $\Pi$, we have to obtain also the
function $ f(x,t) $ of the coordinate transformation.
The function $ f $ can be
evolved with the auxiliary equation
\begin{equation}
\dot f = \beta f' \,\, .
\end{equation}
\end{itemize}
A straightforward von Neumann analysis
shows that the stability of the two sets of
finite differenced equations (\ref{39}), (\ref{40})
and (\ref{41}), (\ref{42})
are the same, as they should be by construction.
Eqns.~(\ref{41}) and (\ref{42}) have the correct
causal property independent
of the value of the shift.

Fig.~\ref{fig:cau} shows the evolution of $\phi$ with this causal leapfrog
differencing scheme using the same initial data, grid
parameters and shift as in Fig.~\ref{fig:cenun}.
The evolution is carried out to $ t = 500 $,
with $ \phi $ shown at equal time intervals.
Clearly the evolution is stable.

\subsection{Causal Differencing in GR}

The situation with the Einstein equations is no different from the
scalar field equation as far as the problem of finite
differencing is concerned.  To find the
finite differenced version
of Eqns.~(\ref{adot}) to (\ref{hbdot}), we transform
to a coordinate system without a shift
\begin{eqnarray}
 \tilde t  = && t     \\
  \tilde \eta = && f(t,\eta ) \,  . \label{45}
\end{eqnarray}
The line element in these new coordinates can be written as
\begin{equation}
   ds^2 = - {\tilde \alpha}^2 d {\tilde t} {}^2
   + \tilde A \, d \tilde\eta^2
   + \tilde B \, (d\theta^2 + \sin^2\theta \, d \phi^2 ) \,  .
\end{equation}
It is easy to see that the two sets of metric functions are
related by
\begin{eqnarray}
  \tilde\alpha      = && \alpha\\
  \tilde A          = && {A\psi^4 \over {f'}^2}\\
  {\dot f\over f'} = && \beta \label{49} \\
  \tilde B          = && B\psi^4 \quad .
\end{eqnarray}
In terms of this new set of variables, the Einstein evolution equations
(\ref{adot}) to (\ref{hbdot}) become
\begin{eqnarray}
     {\partial \tilde A \over \partial \tilde t} \,
= && -2\tilde\alpha \tilde H_A \\
     {\partial \tilde B \over \partial \tilde t} \,
= && -2\tilde\alpha \tilde H_B
\end{eqnarray}
\vspace{-0.1 in}
\begin{eqnarray}
  {\partial\tilde H_A \over \partial\tilde t}
&=& {\tilde \alpha} R_{ \tilde{\eta} \tilde{\eta} }
- {\partial^2\tilde\alpha\over \partial\tilde\eta^2}
+ {1\over 2\tilde A} \,
  {\partial\tilde\alpha \over \partial\tilde\eta} \,
  {\partial\tilde A \over \partial\tilde\eta}
+ 2 {\tilde \alpha} \, {\tilde H_A\tilde H_B \over \tilde B}
- {\tilde \alpha\tilde H_A^2 \over \tilde A} \\
  {\partial\tilde H_B \over \partial\tilde t}
 &=& \tilde\alpha R_{\tilde \theta \tilde \theta}
- {1\over 2\tilde A} \,
  {\partial\tilde\alpha \over \partial\tilde\eta} \,
  {\partial\tilde B\over \partial\tilde\eta}
+ {\tilde\alpha \tilde H_A\tilde H_B \over \tilde A} \,\, .
\end{eqnarray}
Finite differencing these equations is straightforward.
For example, if the
leapfrog scheme is chosen,
upon transforming back to the $ ( t, \eta ) $ coordinates,
the resulting finite differenced equations have exactly
the same structure as (\ref{41}) and (\ref{42}).
There are terms without $ \beta $,
evaluated on the $ n^{\text{th}} $
slice at the causal center of the backward
light cone of the point
$ (n+1,j) $, and terms involving $ \beta $ evaluated on the
$ (n-1)^{\text{th}} $ slice
at the corresponding location of the causal center
on this slice.

Also similar to the scalar field case,
we keep the dependent variables in the
finite differenced equations as
$ ( \tilde A$, $\tilde B$, $\tilde H_A$,
$\tilde H_B$).  They are related to ($A$, $B$, $H_A$, $H_B$) by
the auxiliary function $f$, which is evolved with Eqn.~(\ref{49}).
It is also possible to evolve directly with the variables ($A,$ $B$,
$H_A$, $H_B$) with full causal differencing.
In this case no auxiliary function is needed.  The
details of this approach will be presented elsewhere.

\section{Apparent horizon conditions in a spherical spacetime}
\label{sec:AHBC}

In this section we discuss results obtained by implementing the
horizon boundary scheme.  First we outline the issues common
to all varieties of the boundary scheme, and then we present
results for each different implementation.

To study how well the horizon boundary condition works, we
compare results with those obtained using the BHS black hole code,
which is one of the most accurate codes to date in evolving
spherical black holes.
The most accurate evolutions in the BHS code are obtained with
maximal slicing and zero shift, which is the case to
which we shall be comparing our results.
In this case,
the apparent horizon is initially located on the throat
of the black hole, which is the inner boundary of the
computational domain.  As the spacetime evolves the coordinates
fall towards the hole, so the coordinate location of the horizon
will move out.
The evolution begins with the lapse collapsing to zero behind the
horizon and the radial metric function increasing
as the coordinates collapse inward.
To stop the radial metric function from developing a
sharp peak (as shown in Fig.~\ref{fig:grr}), we smoothly introduce
a horizon locking shift over a period of time.
The phase-in period for the shift typically lasts $\sim 2M$
starting from $ t \sim 1 M $, and so
the full horizon boundary condition is in place by $ t = 3M $,
and the inner grid points close to the singularity
are dropped from the numerical evolution.
We retain 10 to 20 grid points
inside the apparent horizon as buffer zones for added stability.
Having these ``buffer zones'' inside the
horizon helps because of two reasons.  First, in the horizon locking
coordinate the light cones are tilted more
as we go further inside the horizon.
Second, any inaccuracy generated at the innermost grid
point has a longer distance to leak through, and hence decreases
substantially in amplitude, before it can affect the physically
relevant region outside the horizon.

By the time the shift is fully phased in,
the radial metric function is
still of order unity everywhere.  The lapse is also of order unity
throughout the grid, with the smallest value being 0.3 at the
inner boundary of the  grid. The subsequent evolution,
using the same grid parameters as in the BHS code,
gives us a direct comparison between
implementing and not, the horizon boundary condition.

Without the horizon boundary condition, one has to let the lapse
collapse to a value even below $10^{-10}$ in the inner region,
i.e., one does not evolve that part of the spacetime in order to help
prevent the code from crashing.  Nevertheless, the code will still
subsequently
crash as the radial metric function develops a sharp peak
near the horizon (growing
from a value of order 1 to a peak value well over 100 within a span
of just a few grid points, depending on the
resolution).
With the horizon boundary condition in place, there is no need to
collapse the lapse. In fact the lapse is kept frozen from the
time the boundary condition is fully phased in.
We stress that the boundary condition permits accurate evolutions
for extremely long times (see below) with lapses of order
unity throughout the entire calculation in space and in time.

Another issue is the implementation of causal
differencing for the black hole spacetime.  In our first treatment
of this problem~\cite{Seidel92a} we introduced causal differencing in
a first order way by taking into account the direction of light
cones when constructing difference operators,
but without accounting for the width
and the precise location of the center
of the backward light cones.
Basically, the implementation
in \cite{Seidel92a} is similar to using one-sided
derivatives in a region where the flow of information is one-sided.
Here we report on the results obtained with the full causal
differencing scheme using the ``tilde'' variables as
described in Section~IV, and we see a substantial improvement in
accuracy.

We begin by presenting results of implementing the horizon
boundary condition using our ``distance freezing'' shift
that freezes the radial metric function $ A $ in time.
In Fig.~\ref{fig:hlock:a} we show the evolution of the radial metric
function $ A $ up to a time of $ t = 150 M $ for a run with
400 radial zones ($\Delta \eta = 0.015$).
Note that $ \eta $ is a logarithmic radial coordinate
that runs from $\eta=0$ at the throat to $\eta=6$ at the
outer boundary.
The outer edge corresponds to
$ r' = 202 M $ in the isotropic coordinate $ r' $.
The horizon is locked at $ \eta = 0.8 $ after the
boundary scheme is fully phased in by $ t = 3 M $.
The inner grid points are dropped at that point, so the lines
at later times do not cover the inner region.
We see that $ A $
changes rapidly initially
before and during the phase-in period.  After that,
the evolution gradually settles down.
 From $ t = 50 M $ to $ t = 150 M $ the profile barely changes.
Although the shift is designed
to make $A$ a constant in time, due to discretization errors, the
freeze is not perfect.  However, a perfect locking is also not
necessary, as demonstrated by the stability and the accuracy of
the results shown here.
This slow drifting in $A$ is to be
compared to the sharp peak in Fig.~\ref{fig:grr} produced by the BHS
code without using the horizon boundary condition.

In Fig.~\ref{fig:hlock:b}
we show the evolution of $ B $, the angular metric
function.  We see that the shift which is designed to freeze $A$
also freezes $B$.
Again the evolution of $ B $ settles down by $ t = 50 M $,
and the change in the profile becomes negligible.
This shows that the shift vector has succeeded in
locking the coordinate system to the geometry.  With the geometry
of the Schwarzschild spacetime being static, and the time slicing
(lapse) not changing in time, all metric
functions must become frozen.

In Fig.~\ref{fig:hlock:m} we show the mass of the apparent horizon,
defined by
\begin{equation}
\label{ahmass}
M_{AH} = \left. \sqrt{\frac{4 \pi \psi^4 B}{16 \pi}} \,
\right|_{\eta = \eta_{AH} } .
\end{equation}
On the vertical axis we show the difference between
the analytic value $M_{AH}=2$ and the numerically computed value.
By $t=150M$ the mass is just $M_{AH}=2.005$ for the ``distance freezing''
shift with full causal differencing, with resolution
of 200 grid points.
We also show results obtained
with our earlier first order implementation of the boundary
condition scheme, which did not use full causal differencing.
Although the old results are already quite good,
the improvement due solely to causal differencing is
clear.  For comparison, we show also the error in mass given by
the BHS code with the much higher resolution of 400 grid zones.
The improvement one can
get by imposing the horizon boundary condition is obvious.

Figs.~\ref{fig:block:a}, \ref{fig:block:b},
and \ref{fig:block:m},
show, respectively, the evolution of
$A$, $B$ and $M-2$, analogous to
Figs.~\ref{fig:hlock:a}, \ref{fig:hlock:b}, and \ref{fig:hlock:m},
but for the case of the horizon boundary
implemented with the ``area freezing'' shift discussed in Section~IV.
We note that the basic features are the same.
Before and during the phase-in, $ A $ and $ B $ are rapidly changing.
After the phase-in, the metric functions $A$ and
$B$ basically settle down.  Although the shift is designed to
freeze $B$ in this case, $A$ also gets frozen.
We see in Fig.~\ref{fig:block:b} that $B$
is much more accurately locked compared to the $B$
in Fig.~\ref{fig:hlock:b}.
Notice that in Figs.~\ref{fig:block:a} and \ref{fig:block:b},
$A$ and $B$ have different spatial distributions
compared to those of
Figs.~\ref{fig:hlock:a} and \ref{fig:hlock:b}.
The grid points are tied to the geometry
by the shifts in both cases,
but they are tied at different locations.
In the area freezing shift case, the shift at the horizon is treated
as the shift at other positions.  The error in
mass, $M_{AH}-2$ which is a very sensitive indicator of accuracy
is plotted against time for different resolutions
in Fig.~\ref{fig:block:m}.
Again the BHS case is shown for comparison.

The next set of
Figs.~\ref{fig:elock:a}, \ref{fig:elock:b},
\ref{fig:elock:m}, and \ref{fig:expb}
are for the expansion freezing shift,
a choice of shift vector which is geometrically
motivated and coordinate independent.  We again see that the
grid points are basically locked to the geometry by this
condition after phasing in. However, we see that there is an
intermediate region
outside the horizon but not too far from the black hole
for which the locking is less perfect compared
to the previous two choices of shift.
The basic reason is that
although the expansion $\Theta$ is monotonically increasing with
the radial coordinate near the horizon, it is not the case
farther out, as shown in Fig.~\ref{fig:expb}.
Near the peak of the expansion, where the expansion
is nearly a constant with respect to changing radial position,
the determination of a shift required to lock the surface of
constant expansion is clearly more difficult.  As a result,
both $A$ in
Fig.~\ref{fig:elock:a} and $B$ in Fig.~\ref{fig:elock:b}
show some evolution in that region,
indicating motion of the grid.  However, as this
region is further away from the horizon, such motion of the grid
is not causing any serious difficulty, and the evolution is
still stable and accurate, as shown by the error in the mass plot of
Fig.~\ref{fig:elock:m}.

Figs.~\ref{fig:mdlock:a}, \ref{fig:mdlock:b},
and \ref{fig:mdlock:m}
are the corresponding figures
for the final minimal distortion shift case.
BHS reported that the use of a minimal distortion shift is
troublesome in the region near the throat, as the volume
element is small there.  With the horizon boundary condition, the
region near the throat is cut off explicitly, and the minimal
distortion shift with boundary conditions for the shift equation
set on the horizon ({\it cf.} see Sec.~III.B) works as nicely as
the other horizon locking shifts.

As a further check on the accuracy of the calculations, in
Fig.~\ref{fig:alkham}
we show the distributions of the violation of the
Hamiltonian constraint at $ t = 90M $ for the BHS code without a
horizon boundary condition, and at $ t = 150 M $
for the four different implementations
of the horizon conditions.
As the evolutions are completely
unconstrained, the value of the Hamiltonian provides a useful
indicator of the accuracy.
The distributions of the Hamiltonian constraint show
steady profiles in time for all of our
horizon boundary schemes.
We show the comparison to the BHS result at $ t = 90 M $
instead of $ t = 150 M $, since the BHS code
is no longer reliable after $ 90 M $ for
reasons discussed above.
To put the BHS result on the same scale it is
divided by a factor of 10.  All runs are done with the same grid
parameters and a resolution of 400 zones.  The increase in accuracy
provided by the horizon boundary condition is obvious.

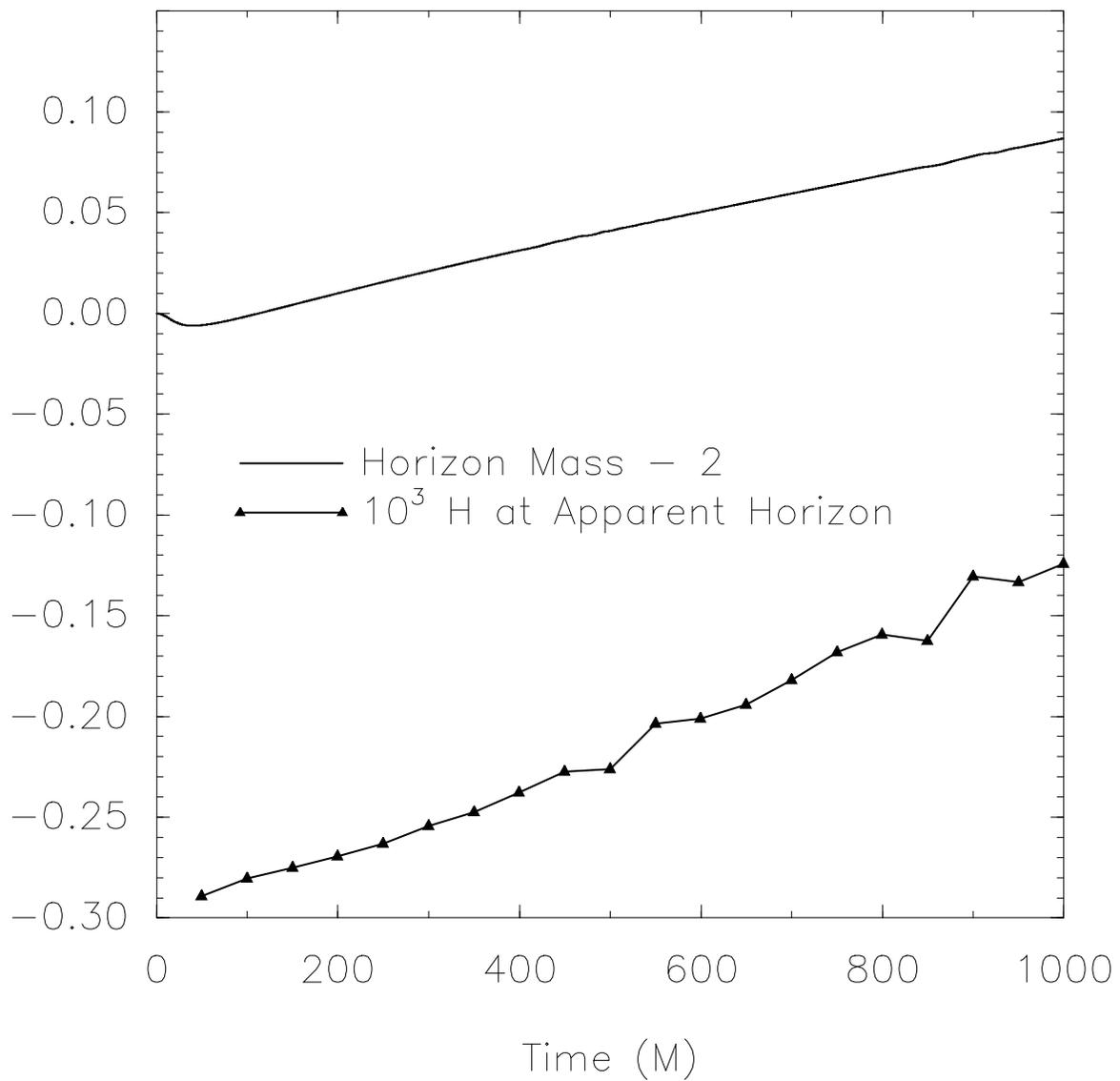
\begin{figure}
\caption{The metric function $ B $  is shown for the case where the
horizon boundary condition is implemented with a
``localized shift''.
Near the apparent horizon of the black hole the shift
is the area freezing shift.
Further out the shift is smoothly lowered to zero,
such that $ \beta = 0 $ for $ \eta > 3.6 $.
In this evolution of one black hole, the grid points
which are not locked to the horizon
fall towards the hole, i.e., $ B $ decreases, as expected.
If additional black holes were present, the motion of the grid points
in the outer region would be different. This demonstrates the
flexibility of the horizon boundary scheme.
\label{fig:loclk}
}
\end{figure}

For spacetimes with multiple black holes,
the grid points should not all be locked with respect to
just the horizon of one of the holes.
In Fig.~\ref{fig:loclk} we show the effect of turning off
the shift at a finite distance  away from the black hole.
The shift is smoothly set to zero at $ \eta = 3.6 $, with the shift
at small $\eta$
being the area freezing shift. In the region that the coordinate is not
locked,
we see in Fig.~\ref{fig:loclk} that the value
of $ B $ is decreasing in time as expected,
since the grid points there are falling towards the hole.
The motion of the grid points would be different if there
were another hole further out in the spacetime.
This study of using a ``localized shift''
provides evidence that our scheme is flexible enough to
handle more complicated situations.

One last question one might have is the long term stability of
the horizon boundary scheme.
In Fig.~\ref{fig:lgtrm}, we show a run up to
$t=1000M$ with a resolution of 200 grid zones
and using the ``distance freezing''
shift.  Errors in the mass $ M_{AH} - 2 $
and the Hamiltonian constraint evaluated at the horizon are shown
as a function of time.
We note that at the end of the run the mass error is just 4\%.
Stable and accurate evolutions for such a long time are likely to
be required for the spiraling two black hole coalescence
problem.

\section{discussion}
\label{sec:discussion}

Progress in numerical relativity has been hindered for 30 years
because of the difficulties in avoiding spacetime singularities
in the calculations.  In this paper,
we have presented several working examples of how
an apparent horizon boundary scheme can help
circumvent these difficulties.
We have demonstrated this scheme
to be robust enough
that it allows many different ways of implementation.
Also, as shown by the results in Section V, it is
likely that even an approximate implementation of horizon
boundaries can
yield stable and accurate evolutions of black hole spacetimes.
Such approximate implementations can be most important in extending
this work to 3D spacetimes and will be discussed elsewhere.
Throughout this work we have sought to consider ideas that are
applicable or extensible to the most
general 3D case, where no particular symmetry, gauge condition, metric form,
or evolution
scheme is assumed.  In fact four different gauge conditions were demonstrated
here with the same code, and in no case were the equations specialized
to the particular gauge under investigation.

Although in this paper we have only implemented
this scheme in spherical vacuum spacetimes,
neither sphericity nor vacuum are intrinsic
restrictions on the scheme.
Indeed, in~\cite{Seidel92a}
we have shown that the horizon boundary condition scheme
works in the case of a scalar field infalling into
a black hole. The follow-up study of that case will be
reported elsewhere.

The discussion in this paper has been carried out in terms of
a free evolution scheme. For constrained evolution,
the horizon boundary condition can be constructed in
the same spirit as discussed here.
With the implementation of a horizon locking shift,
the boundary values of the elliptic constraint equations
can be obtained from data
in the region of causal dependence on the previous
time slice, using the evolution equations.  Furthermore, the apparent
horizon condition itself could be used to provide a boundary value
of some quantity for constraint equations.

Although we have shown only cases where the horizon
is locked to the coordinate system, as pointed out in Section III,
this is not a requirement
for the evolution.
The shift can be used to control the motion of the horizon.
Such controlled motion will be particularly useful in
moving black holes through the numerical grid, as will
be needed, for example, in the long term evolution
of the two black hole inspiral coalescence.
We found that such controlled motion can be easily achieved
in the scheme described in this paper. The details of
this will be reported elsewhere.

Other issues that need to be addressed are that the
apparent horizon location may jump
discontinuously, or multiple horizons may form.
These problems can be handled by simply tracking newly formed horizons
and phasing in a new boundary in place of the old one(s).
Multiple black holes can be handled by locking the coordinates
only in the vicinity of the black holes.
We are beginning to examine these issues.

\section*{acknowledgements}

We are happy to acknowledge helpful discussions with Andrew Abrahams,
David Bernstein, Matt Choptuik, David Hobill, Ian Redmount, Larry
Smarr, Jim Stone, Kip Thorne, Lou Wicker, and Clifford Will.  We are
very grateful to David Bernstein for providing a copy of his black
hole code, on which we based much of this work.  J.M. acknowledges a
Fellowship (P.F.P.I.) from Ministerio de Educaci\'on y Ciencia of
Spain. This research is supported by the NCSA, the Pittsburgh
Supercomputing Center, and NSF grants Nos. PHY91-16682, PHY94-04788,
PHY94-07882 and PHY/ASC93-18152 (arpa supplemented).


\begin{figure}
\caption{A black hole spacetime diagram showing various singularity
avoiding time slices that wrap up around the singularity inside
the horizon.  Such slicings allow short term success in the numerical
evolution of black holes, while at the same time causing
pathological behavior
that eventually dooms the calculation at late times.
\label{fig:wrap}
}
\end{figure}

\begin{figure}
\caption{The radial metric function $A$ is shown at intervals
of $ 10 M $ up to $ t = 100 M $ for the black hole evolution
using a code developed by BHS.  The location of
the apparent horizon  is shown on each
time slice as a dotted line. As time increases,
the height of the peak grows without bound.
The calculation was performed
with 400 radial zones ($ \Delta \eta = 0.015 $)
with maximal slicing and zero shift, the best case
reported in BHS for this resolution.
\label{fig:grr}
}
\end{figure}

\begin{figure}
\caption{The Hamiltonian constraint is shown at
intervals of $ 10 M $ up to $ t = 100 M $
for the BHS black hole evolution
with maximal slicing and zero shift with 400 grid zones.
As the sharp peak develops in the radial metric function,
the violation of the
Hamiltonian constraint becomes quite large. Its peak value reaches
a maximum near time $ t \sim 90 M $ when the resolution
is no longer sufficient to resolve the peak in
the radial metric function $ A $.
\label{fig:ham}
}
\end{figure}

\begin{figure}
\caption{The evolution of a scalar field $ \phi $ using the
standard leapfrog scheme is shown at equally spaced
intervals up to time $ t = 210 $ with the shift
$ \beta (x, t) $
specified by Eq. (37).
In the case of zero shift
the initial gaussian splits into two components
which propagate in opposite directions.
Employing the shift in Eq. (37)
causes the `left traveling' component to have
a large coordinate velocity to the left, while the
`right traveling' component has a smaller coordinate velocity.
The use of the standard leapfrog scheme leads to an instability
which arises when the ``region of finite differencing''
fails to cover the ``region of causal dependence.''
\label{fig:cenun}
}
\end{figure}

\begin{figure}
\caption{The evolution of a scalar field $ \phi $ using the
causal leapfrog differencing scheme (Eqs.~(40) and (41))
is shown at equally spaced
intervals up to time $ t = 500 $ with the same coordinate parameters
and shift as in Fig. 4.
The use of causal differencing leads to stable evolutions
independent of the value of the shift. The finite differencing
scheme follows the causal structure of the coordinate grid
and covers the ``region of causal dependence.''
\label{fig:cau}
}
\end{figure}

\begin{figure}
\caption{The evolution of the radial metric function $ A $
for the ``distance freezing'' shift
is shown at a resolution of 400 grid zones.
Due to the use of this shift, the large peak that develops
in this function in the zero shift case has been completely eliminated
and the evolution proceeds free of its associated difficulties.
Notice that after the phase-in of the horizon boundary condition
at $t\sim 3M$, the inner-most grid points are dropped from
the computational domain. The metric function gradually settles
down to a steady profile.
\label{fig:hlock:a}
}
\end{figure}

\begin{figure}
\caption{The evolution of the angular metric function $ B $
for the ``distance freezing'' shift
is shown for the same run as in Fig. 6.
The function $ B $ gradually settles to steady profile
after the implementation of the horizon boundary condition.
\label{fig:hlock:b}
}
\end{figure}

\begin{figure}
\caption{The error in the horizon mass is shown
for the ``distance freezing'' shift as applied in both the
present causal differencing scheme and the first order
method of [18]
for runs made
with a resolution of 200 grid zones.
The BHS result for maximal slicing and zero shift
at a resolution of 400 grid zones is also shown.
The difference between the computed and analytic value of 2 is
plotted against time.  Note the difference in the slopes
of the curves.
\label{fig:hlock:m}}
\end{figure}


\begin{figure}
\caption{The evolution of the radial metric function $ A $
for the ``area freezing'' shift is shown
for a resolution of 400 grid zones.
Although the shift is designed to lock $ B $,
large peaks in the function $ A $
are also prevented by the use of this
shift. There is an interval of time during and
shortly after the phase-in
of the horizon boundary condition when the function $ A $ drifts,
turning downward at the inner edge.  By the time the run
reaches $ 50 M $ the curve is steady.
Note also the dropping of the innermost points from the computational
domain after the phase-in period.
\label{fig:block:a}
}
\end{figure}

\begin{figure}
\caption{The angular metric function $ B $
for the ``area freezing'' shift is plotted against the radial
coordinate $ \eta $ on many time slices for the same run as Fig. 9.
The function $ B $ is locked extremely well with this choice of shift.
\label{fig:block:b}
}
\end{figure}

\begin{figure}
\caption{The error in the horizon mass obtained in using the
horizon boundary condition with
the ``area freezing'' shift at two resolutions. Also shown is the
BHS result for maximal slicing
and zero shift with 400 grid zones.
The difference between the computed and analytic value of 2
is plotted against time.  Note the difference in the slopes
of the curves.
\label{fig:block:m}
}
\end{figure}

\begin{figure}
\caption{The evolution of the radial metric function $ A $
for the ``expansion freezing'' shift
is shown at a resolution of 400 grid zones.
The function $ A $ is again free of sharp peaks for this
choice of shift. A region of space near the maximum of
$ \Theta $ shows some evolution,
i.e., some motion of the coordinates, but this does not cause
any loss of accuracy or stability, in contrast to the
development of a sharp peak.
\label{fig:elock:a}
}
\end{figure}

\begin{figure}
\caption{The evolution of the angular metric function $ B $
for the ``expansion freezing'' shift is shown at a resolution of 400 grid
zones. The function $ B $ shows some motion of the coordinate grid
in the region where $ \Theta $ is near its maximum.
\label{fig:elock:b}
}
\end{figure}

\begin{figure}
\caption{The error in the horizon mass
for the ``expansion freezing'' shift for two resolutions is shown
along with the BHS
result for maximal slicing and zero shift
at 400 grid zone resolution.
We plot the
difference between the computed and analytic value of 2 against time.
\label{fig:elock:m}
}
\end{figure}

\begin{figure}
\caption{The evolution of the function $ B \Theta $
for the ``expansion freezing'' shift
is shown for a run of $ 150 M $ at a resolution of
400 grid zones. This choice of shift is designed
to freeze $ B \Theta $ and the resulting profile is very steady,
as soon as the the horizon boundary condition is phased in.
\label{fig:expb}
}
\end{figure}

\begin{figure}
\caption{The evolution of the radial metric function $ A $ is shown
for the minimal distortion shift at a resolution of 400 grid zones.
The horizon boundary condition is phased in from $t\sim 1M$ to
$t\sim 3M$, during which the inner-most grid points are dropped.
The function $ A $ is again free of sharp peaks for this
choice of shift and is locked in place well after $t\sim 50M$.
\label{fig:mdlock:a}
}
\end{figure}

\begin{figure}
\caption{The angular metric function $ B $
for the minimal distortion shift is plotted against radius $ \eta $
for the same run as Fig. 16. After it settles down at $t\sim 50M$,
the function $ B $ maintains a nearly constant profile over the
$ 150 M $ run.
\label{fig:mdlock:b}
}
\end{figure}

\begin{figure}
\caption{The error in the horizon mass
for the minimal distortion shift is shown at two resolutions
along with the BHS
result for maximal slicing and zero shift at
a resolution of 400 grid zones.
The difference between the computed and analytic value of 2 is plotted
against time.
\label{fig:mdlock:m}
}
\end{figure}

\begin{figure}
\caption{The violation of the Hamiltonian constraint is shown
at time $ t = 150 M $ for several different implementations of the
horizon boundary condition. The curves of the Hamiltonian
shown here for all of our choices of shift
vector are nearly constant in time once the phase-in of
the horizon boundary scheme is complete.
For comparison we show the Hamiltonian
constraint for the BHS code, divided by a factor of 10,
at time $ t = 90 M $, since after this time the BHS results at this
resolution are unreliable.
All runs are made at the same
resolution of 400 grid zones.
\label{fig:alkham}
}
\end{figure}

\begin{figure}
\caption{The Hamiltonian constraint at the
apparent horizon and the horizon mass
for the horizon boundary condition using the
``distance freezing'' shift are shown
for a run up to $t = 1000 M $ with 200 grid zone resolution.
The error in the mass increases slowly and in a
linear fashion over the majority of the run.
The code can run to even longer times
with this same slow rate of increase in
the mass error.
\label{fig:lgtrm}
}
\end{figure}

\end{document}